\documentstyle[pra,aps,twocolumn,epsf,amssymb]{revtex}

\begin{document}
\draft
\title{Bell State Preparation using Pulsed Non-Degenerate Two-Photon Entanglement}
\author{Yoon-Ho Kim,\thanks{Email: yokim@umbc.edu} Sergei P. Kulik,\thanks{Permanent address:
Department of Physics, Moscow State University, Moscow, 119899,
Russia.} and Yanhua Shih}
\address{Department of Physics, University of Maryland, Baltimore
County, Baltimore, Maryland 21250}
\date{Revised, 9 October, 2000}

\maketitle

\vspace*{-12mm}

\widetext

\begin{abstract}
We report a novel Bell state preparation experiment. High-purity
Bell states are prepared by using femtosecond pulse pumped
\emph{nondegenerate} collinear spontaneous parametric
down-conversion. The use of femtosecond pump pulse {\em does not}
result in reduction of quantum interference visibility in our
scheme in which post-selection of amplitudes and other traditional
mechanisms, such as, using thin nonlinear crystals or narrow-band
spectral filters are not used. Another distinct feature of this
scheme is that the pump, the signal, and the idler wavelengths
are all distinguishable, which is very useful for quantum
communications.
\end{abstract}


\narrowtext

\vspace*{-5mm}

Preparation and measurement of the Bell states are two important
issues in modern quantum optics, especially for quantum
communications, quantum teleportation, etc \cite{bellstates}. For
photons, such states can be realized by using the entangled
photon pairs generated in spontaneous parametric down-conversion
(SPDC). By making appropriate local operations on the SPDC photon
pairs, one can prepare all four Bell states.

The polarization Bell states, for photons, can be written as
\begin{eqnarray}
|\Phi^\pm\rangle&=&|X_1,X_2\rangle\pm|Y_1,Y_2\rangle,\nonumber\\
|\Psi^\pm\rangle&=&|X_1,Y_2\rangle\pm|Y_1,X_2\rangle,
\label{bellstates}
\end{eqnarray}
where the subscripts 1 and 2 refer to two different photons,
photon 1 and photon 2, respectively, and they can be arbitrarily
far apart from each other. $|X\rangle$ and $|Y\rangle$ form the
orthogonal basis for the polarization states of a photon, for
example, it can be horizontal ( $|H\rangle$) and vertical
($|V\rangle$) polarization state, as well as $|45^\circ\rangle$
and $|-45^\circ\rangle$, respectively. This means that the
quantum interference should be independent of the choice of the
bases.

Such an experiment was first performed by Shih and Alley in which
non-collinear type-I SPDC and a beamsplitter were used to prepare
a Bell state \cite{shih}, but it is very difficult to align such a
system. Collinear type-II SPDC is thus developed \cite{Kiess}.
There is, however, a common problem: the entangled photon pairs
have 50\% chances of leaving at the same output ports of the
beamsplitter. Therefore, the state prepared after the
beamsplitter may not be considered as a Bell state without
amplitude post-selection \cite{garuccio}. Only when one considers
the coincidence contributing terms by throwing away two out of
four amplitudes (post-selection of 50\% of the amplitudes), the
state is then said to be a Bell state. This problem is later
 solved by using  non-collinear
type-II SPDC  or using two non-collinear type-I SPDC
\cite{kwiat1}.

 \newpage \vspace*{24mm} In the cw pumped SPDC, entangled photon pairs occur randomly
since the process is ``spontaneous", so whereabouts of the photon
pair is completely uncertain within the coherence length of the
pump laser beam. This huge time uncertainty makes it difficult
for applications such as generation of multi-photon entangled
state, quantum teleportation, etc, as interactions between
entangled photon pairs generated from different sources are
required. This difficulty was thought to be solved by using a
femtosecond pulse laser as a pump. Unfortunately, femtosecond
pulse pumped type-II SPDC shows poor quantum interference
visibility due to the very different (compared to the cw case)
behavior of the two-photon effective wave-function \cite{keller}.
One has to utilize special experimental schemes to achieve
complete overlap of the two-photon amplitudes. Traditionally, the
following methods were used to restore the quantum interference
visibility in femtosecond pulse pumped type-II SPDC: (i) use a
thin nonlinear crystal ($\approx100 \mu$m) \cite{sergienko} or
(ii) use narrow-band spectral filters in front of detectors
\cite{keller,femto}. Both methods, however, reduce the available
flux of the entangled photon pair significantly \cite{critic} and
cannot achieve complete overlap of the wave-functions in principle
\cite{keller}.

The first attempt to achieve high-visibility quantum interference
in femtosecond pulse pumped type-II SPDC without using narrowband
filters and a thin crystal was reported in Ref.\cite{branning}.
The observed visibility, however, was rather low and keeping the
phase coherence over a long term would be very difficult since a
Michelson interferometer is used. Also, such a scheme cannot be
used to prepare a Bell state. Recently, we reported a
high-visibility quantum interference experiment in which photon
pairs are entangled both in polarization and space-time using
femtosecond pulse pumped type-I SPDC \cite{kim1}. However, it
cannot be considered as a true Bell state preparation since
post-selecting 50\% of the amplitudes was still necessary.

In this Letter, we report a Bell state preparation experiment in
which we effectively eliminate \emph{any post-selection} in
femtosecond pulse pumped SPDC for the first time. Other features
in our scheme include: (i) collinear SPDC makes the alignment
much easier, (ii) Alice and Bob shares photons of different
frequencies entangled in both space-time and polarization, (iii)
phase coherence is automatically kept and the visibility as high
as 92\% is observed, (iv) thick crystals can be used to increase
the intensity (without losing the visibility), and (v) the
spectral bandwidth is reduced significantly by the use of
\emph{nondegenerate} SPDC. These features make our scheme a good
source of Bell states for quantum information experiments.

\begin{figure}[tbp]
\centerline{\epsfxsize=3in\epsffile{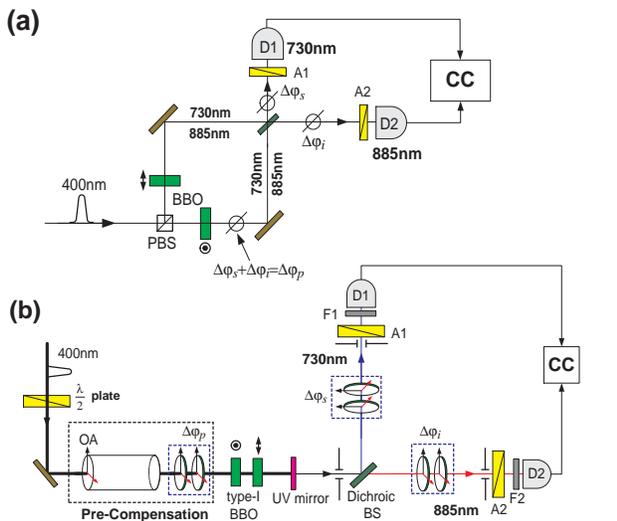}}\caption{(a)
Principle schematic of the experiment. The pump pulse is
polarized in $45^\circ$. Non-degenerate collinear type-I SPDC
occurs at the nonlinear crystal placed in each arm of the MZI.
(b) Schematic of the experimental setup. Note that three different
phases can be observed. Interference filters $F_1$ and $F_2$ are
used to cut the pump noise. }\label{fig:setup}
\end{figure}

The basic idea of the experiment is illustrated in
Fig.\ref{fig:setup}(a). A $45^\circ$ polarized femtosecond laser
pulse (central wavelength $\lambda_p=400$nm and pulse duration
$\sigma=80$fsec.) enters the Mach-Zehnder interferometer (MZI)
which contains a type-I nonlinear crystal in each arm. One
crystal has its optic axis oriented vertically ($\updownarrow$)
and another horizontally ($\odot$). Polarizing beamsplitter (PBS)
splits the $45^\circ$ polarized pump pulse into the vertical and
horizontal polarized pulses propagating along different arms of
the MZI. Then non-degenerate collinear type-I SPDC occurs, with
equal probability, at each crystal (signal wavelength $=730$nm and
idler wavelength $=885$nm) and they are mixed at the dichroic
beamsplitter which directs 730nm photons to detector $D_1$ and
885nm to detector $D_2$. In the simplified single mode
approximation, the quantum state generated from the vertically
oriented crystal ($\updownarrow$) is
$|\psi\rangle_1=|H_{730}\rangle|H_{885}\rangle$ and from the
horizontally oriented one ($\odot$)
 is $|\psi\rangle_2=|V_{730}\rangle|V_{885}\rangle$. $H$ and $V$ represent
horizontal and vertical polarization state of a single photon
respectively. Subscripts 730 and 885 refer to the wavelengths
730nm and 885nm, respectively. When the MZI is balanced, the
quantum state after the MZI is ({\em without} throwing away any
amplitudes)
\begin{equation}
|\Phi\rangle=|V_{730}\rangle_1|V_{885}\rangle_2 +
e^{i\Delta\varphi}|H_{730}\rangle_1|H_{885}\rangle_2,\label{state1}
\end{equation}
where $\Delta\varphi$ is the relative phase between the two
amplitudes and it can easily be varied by scanning one of the
mirror of the MZI.

The coincidence counting rate ($R_c$) is calculated as
\cite{glauber,klyshko},
\begin{equation}
R_c=\int\int dt_+ dt_{12}\left|{\mathcal{A}}(t_+,t_{12})\right|^2,
\label{coinc}
\end{equation}
where $t_{12}=t_1-t_2$ and $t_+=(t_1+t_2)/2$. $t_i=T_i-l_i/c$
where $T_i$ is the time at which detector $i$ fires and $l_i$ is
the optical path length from the surface of the crystal to the
detector $i$. ${\mathcal{A}}(t_+,t_{12})$ is the amplitude of the
biphoton as explicitly calculated in Ref.\cite{keller}.

For the scheme shown in Fig.\ref{fig:setup}(a),
${\mathcal{A}}(t_+,t_{12})$ is the sum of the two amplitudes
originated from the crystal in each arm of the MZI:
\begin{equation}
{\mathcal{A}}(t_+,t_{12})={\mathcal{A}}_a(t_+,t_{12})+{\mathcal{A}}_b(t_+,t_{12}),
\end{equation}
where the subscripts $a$ and $b$ refer to the crystal from which
the amplitudes are created. The delay $T$ introduced in one arm
modifies the amplitude
${\mathcal{A}}_b(t_+,t_{12})\rightarrow{\mathcal{A}}_b(t_++T,t_{12})$
 and determines the additional phase shift for the biphoton amplitudes
$\Delta\varphi=\Omega_p T=K_p \Delta x$, where
$K_p=2\pi/\Lambda_p$, $\Omega_p(\Lambda_p)$ the central frequency
(wavelength) of the pump, and $\Delta x$ the spatial delay. Due to
the energy conservation and negligibly small dispersion of the
air, the phase shift depends only on the  pump wavelength $\Delta
\varphi=K_s\Delta x + K_i \Delta x = K_p \Delta x = \Delta
\varphi_p$ although the delay is introduced to the SPDC field
\cite{kim1,burlakov}. If the crystals are the same and the pump
fields in different arms of the MZI are identical,
\begin{equation}
|{\mathcal{A}}_a(t_+,t_{12})|=|{\mathcal{A}}_b(t_+,t_{12})|.
\end{equation}
The coincidence counting rate is then calculated to be
\begin{equation}
R_c=1+V\cos(\Omega_p T),\label{Rc}
\end{equation}
where $V\approx 1$ in this experiment \cite{note1}. Note that the
angles of the analyzers $A_1$ and $A_2$ are assumed to be
$45^\circ$. From Eq.(\ref{Rc}), we expect that the coincidence
counting rate will be modulated in the pump central wavelength
when $T$ is varied.

There are also two more ways to vary the phases of interference by
introducing relative delays (using a piece of birefringent
material, such as a quartz plate) after the output beamsplitter,
i.e., in the signal ($\Delta\varphi_s$) and/or in the idler
($\Delta\varphi_i$) channels. Therefore we obtain
\begin{equation}
R_c=1+V\cos(\Delta\varphi_p-\Delta\varphi_i-\Delta\varphi_s)
,\label{Rc2}
\end{equation}
where $\Delta\varphi_p$, $\Delta\varphi_i$, and $\Delta\varphi_s$
refer to the relative phases of the pump, the idler, and the
signal, respectively.

As we have shown so far, one can eliminate the possibility of the
entangled photon pairs leaving at the same output ports of the
beamsplitter by employing non-degenerate two-photon entanglement.
In this scheme, high-visibility quantum interference can be
achieved independent of the crystal thickness and the spectral
filter bandwidths even with femtosecond pulse pump.

In practice, however, one would not like to use a MZI in the
experimental setup due to stability related issues. Therefore, we
use a collinear scheme where two type-I BBO crystals are placed
collinearly in the pump beam path, see Fig\ref{fig:setup}(b). Two
type-I BBO crystals with thickness 3.4mm each (the first one is
oriented horizontally and the second one is oriented vertically)
are then pumped by a $45^\circ$ polarized pump pulse. As
described before, the quantum state resulting from the first BBO
is $|V_{730}\rangle|V_{885}\rangle$ and that from the second BBO
is $|H_{730}\rangle|H_{885}\rangle$. Since both crystals are
pumped equally, the two amplitudes are equally probable. Due to
the dispersion, however, $|V_{730}\rangle|V_{885}\rangle$ from
the first BBO ($\odot$) and $|H_{730}\rangle|H_{885}\rangle$ from
the second BBO ($\updownarrow$) are distinguishable in time. To
make $|V_{730}\rangle|V_{885}\rangle$ and
$|H_{730}\rangle|H_{885}\rangle$ indistinguishable in time, one
needs to compensate the delay experienced by the SPDC photon
pairs at each crystal. This compensation can be made by using a
properly oriented quartz rod. If the compensation is made
properly, either before or after the down-conversion non-linear
crystals, one will observe high-visibility quantum interference
without any spectral post-selection.

In the collinear scheme, having a perfect temporal compensation
is difficult when the signal wavelength differs very much from
the idler wavelength. This is because the signal-idler photon
pairs created from the first BBO ($\odot$) experience different
dispersion when they pass through the second BBO
($\updownarrow$). (The MZI scheme does not have this
disadvantage). In this experiment, for the wavelengths we are
interested in, the temporal separation is rather small and it
does not affect the interference visibility. To prevent further
dispersion effects, the compensation is made before the BBO
crystals. The compensator consists of a quartz rod and two quartz
plates whose optic axes are oriented vertically, see
Ref.\cite{kim1}, and it imposes roughly 1.5psec required delay
between the $H$- and $V$-polarized 400nm pump pulse which is
mainly determined by the thickness of the BBO crystals. By
tilting the two quartz plates in the opposite directions, the
phase delay $\Delta\varphi_p$ can be varied to prepare a Bell
state. After the two BBO crystals, the remaining UV radiation is
blocked by a UV reflecting mirror and the collinear SPDC is
selected by a diaphragm. Then a dichroic beamsplitter is used to
reflect the signal (730nm) to $D_1$ and to transmit the idler
(885nm) to $D_2$. Two quartz plates are inserted in each beam
path to vary the relative phase of the signal or the idler
independently. The detector package consists of a single-photon
counting module, an interference filter which is used to cut the
pump noise \cite{filter}, and a polarization analyzer.

\begin{figure}[tbp]
\centerline{\epsfxsize=3in\epsffile{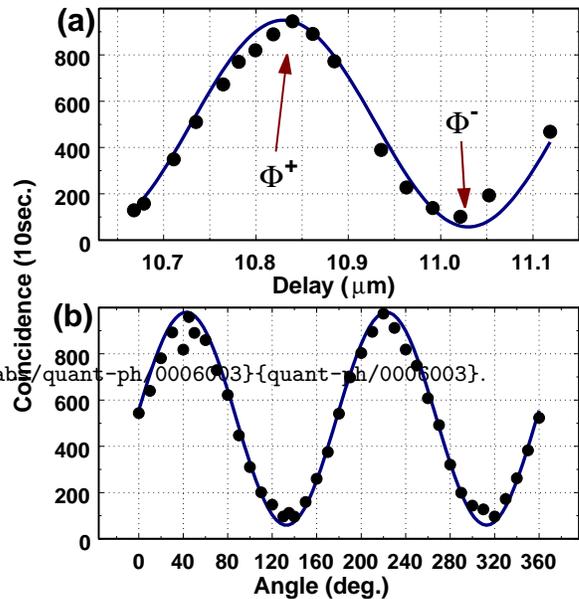}}\caption{Experimental
data. (a) Space-time interference by varying the pump phase
$\Delta\varphi_p$ when $\theta_1=\theta_2=45^\circ$. Transition
from $\Phi^+$ to $\Phi^-$ is clearly demonstrated. (b)
Polarization interference at $\Phi^+$. $A_1$ is fixed at
$\theta_1=45^\circ$ and $A_2$ is rotated, i.e., $\theta_2$ is
varied. Solid lines are the theoretical curves. }\label{fig:data1}
\end{figure}

To demonstrate the effectiveness of this scheme, we first study
the space-time interference as a function of $\Delta\varphi_p$ by
setting $\theta_1=\theta_2=45^\circ$, where $\theta_1$ and
$\theta_2$ are the angles of the analyzers $A_1$ and $A_2$
(measured from the vertical direction). According to
Eq.(\ref{Rc2}), one should observe a pump wavelength modulation
in the coincidence counting rate. Note that $\Delta\varphi_s$ and
$\Delta\varphi_i$ are fixed. The observed modulation period is
400nm, see Fig.\ref{fig:data1}(a), which agrees with the theory.

To prepare $\Phi^+$ ($\Phi^-$) state, identified by constructive
(destructive) interference, one just needs to set
$\Delta\varphi_p-\Delta\varphi_s-\Delta\varphi_i=0, 2\pi, 4\pi,
\ldots$ ($\Delta\varphi_p-\Delta\varphi_s-\Delta\varphi_i=\pi,
3\pi, 5\pi, \ldots$) which can be done by tilting the quartz
plates so that the space-time interference fringe is at the
maximum (minimum). Note that $\Psi^+$ and $\Psi^-$ Bell states
can also be easily prepared by introducing a $\lambda/2$ plate in
one output port of the dichroic beamsplitter.

We have also experimentally demonstrated the polarization
interference for $\Phi^+$ and $\Phi^-$. For $\Phi^\pm$, the
coincidence counting rate is calculated to be
\begin{equation}
R_c\propto\left|\langle\theta_2,\theta_1|\Phi^\pm\rangle\right|^2\propto\cos^2(\theta_1\mp\theta_2).
\label{bellstate1}
\end{equation}
This means that one should observe high-visibility modulation in
polarization correlation measurement for {\em arbitrary values} of
$\theta_1$ and $\theta_2$. To confirm this experimentally, we
first set $\theta_1=45^\circ$ and varied $\theta_2$.
High-visibility polarization correlation is observed, see
Fig.\ref{fig:data1}(b). We then repeated this measurement for many
other values of $\theta_1$ and observed that the visibility
remained the same. This confirms Eq.(\ref{bellstate1}). In other
words, we have successfully prepared polarization Bell states
from femtosecond pulse-pumped SPDC without amplitude and spectral
post-selection.

\begin{figure}[tbp]
\centerline{\epsfxsize=3in\epsffile{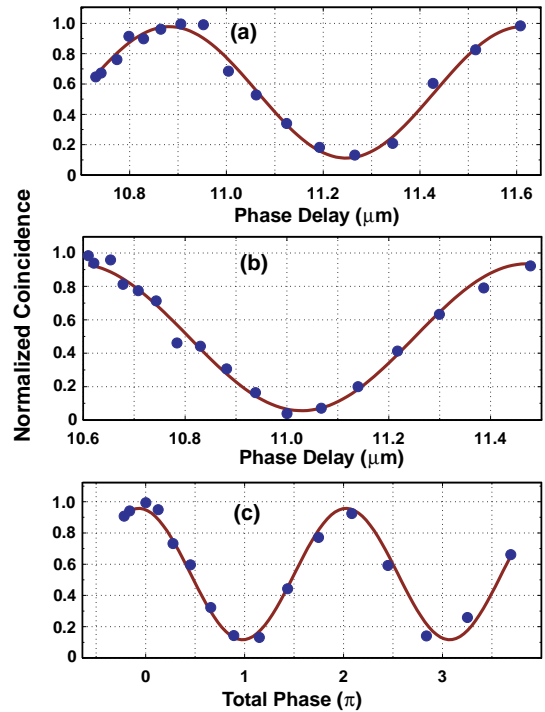}}\caption{Experimental
data. (a) Space-time interference by varying $\Delta\varphi_s$
(730nm modulation). (b) Space-time interference by varying
$\Delta\varphi_i$ (885nm modulation). (c) Two phases (730nm and
885nm) are varied at the same time. Solid line is the theory
curve based on 400nm modulation and agrees well with the data.
This confirms $\Delta\varphi_s+\Delta\varphi_i=\Delta\varphi_p$.
Note that $\theta_1=\theta_2=45^\circ$ for all cases.
}\label{fig:data2}
\end{figure}

Unlike the usual degenerate two-photon sources, this source has
one distinctive feature: one can vary three different phases
independently which is very useful for quantum communications.
To demonstrate this interesting feature, we observe the
space-time interference by varying the relative phases of the
signal ($\Delta\varphi_s$) and the idler ($\Delta\varphi_i$)
independently. In these measurements, $\theta_1$ and $\theta_2$
are set at $45^\circ$. The effect of $\Delta\varphi_p$ is already
demonstrated in Fig.\ref{fig:data1}(a). When the signal phase
$\Delta\varphi_s$ is varied, see Fig.\ref{fig:data2}(a), the
signal wavelength (730nm) modulation is observed in coincidence
rate, while varying the idler phase $\Delta\varphi_i$, see
Fig.\ref{fig:data2}(b), the idler wavelength (885nm) modulation
is observed. Fig. \ref{fig:data1}(a), Fig.\ref{fig:data2}(a), and
Fig.\ref{fig:data2}(b) clearly demonstrate Eq.(\ref{Rc2}).

Finally, we tested the condition
$\Delta\varphi_p=\Delta\varphi_s+\Delta\varphi_i$ by varying the
signal phase $\Delta\varphi_s$ and the idler phase
$\Delta\varphi_i$ at the same time. The quartz plates in both the
signal and the idler paths are tilted at the same time with equal
angles. To see whether the data agrees with the theory, total
phases accumulated in both beam path are calculated from the tilt
angle, i.e., $\Delta\varphi_s+\Delta\varphi_i$. As evidenced from
Fig.\ref{fig:data2}(c), the data agrees well with Eq.(\ref{Rc2}).

In summary, we have demonstrated a novel scheme to prepare pulsed
entangled photon pairs from which all four Bell states can be
easily obtained. Amplitude and spectral post-selection are not
necessary. Note also that non-maximally entangled states can be
prepared by changing the relative intensities of the pump beams.
The visibility and the photon flux is greatly enhanced by this
method although a femtosecond pulse laser is used as a pump. The
signal, the idler, and the pump phases can be varied
independently with different modulation frequency which is very
useful for quantum communications.

We would like to thank M.H. Rubin for helpful discussions. This
research was supported, in part, by the Office of Naval Research,
ARDA and the National Security Agency. S.P.K. also thanks the
Russian Fund for Fundamental Research Grant No. 99-02-16419 for
partially supporting his visit to UMBC.

\vspace*{-5mm}

\vspace*{-5mm}

\vspace*{-5mm}

\begin{references}
\bibitem{note1} The visibility factor $V$ is in general a function
of $D_+=1/u_0(\Omega_p/2)-1/u_0(\Omega_p)$,
$D''=d^2K_0/d\Omega^2(\Omega=\Omega_p/2)$, $u_0$ the group
velocities of the ordinary ray, $\sigma$ the pump pulse duration,
and the delay $T$. Note that $V$ is not a function of bandwidths
of the spectral filters. One can also prepare a Bell state using
two type-II SPDC. Eq.(\ref{Rc2}) still remains valid, but the
function $V$ has different width and shape. We will discuss the
details of the theory and experiments of this scheme by using
type-I and type-II SPDC in \cite{prep}.
\bibitem{filter} The bandwidth of the spectral filter is 10nm
at full width half maximum. This is comparable to the spectrum
bandwidth of type-I nondegenerate SPDC which is $\approx$ 13nm
for the crystals used in the experiment.
\end{references}
\end{document}